\documentclass[11pt]{emulateapj}\usepackage{times}

\usepackage{natbib,rotating,graphicx}
\usepackage[usenames]{color}
\def\1023{PSR J1023+0038}
\def\xss{XSS J12270-4859}
\def\m28{IGR J18245--2452}

\begin{document}

\title{A propeller model for the sub-luminous disk state of the transitional millisecond pulsar {\1023}}
 \author{A. Papitto\altaffilmark{1}, \& D.~F.~Torres\altaffilmark{1,2}}
  \altaffiltext{1} {Institute of Space
  Sciences (IEEC-CSIC), Campus UAB, Carrer de Can Magrans S/N, 08193 Cerdanyola del Vall\`es, Barcelona, Spain  }
  \altaffiltext{2} {Instituci\'o
  Catalana de Recerca i Estudis Avan\c{c}ats (ICREA), 08010 Barcelona,
  Spain }

\begin{abstract}

The discovery of millisecond pulsars switching between states powered
either by the rotation of their magnetic field or by the accretion of
matter, has recently proved the tight link shared by millisecond radio
pulsars and neutron stars in low-mass X-ray binaries. Transitional
millisecond pulsars also show an enigmatic intermediate state in which
the neutron star is surrounded by an accretion disk, it emits
  coherent X-ray pulsations, but is sub-luminous in X-rays with
respect to accreting neutron stars, and is brighter in gamma-rays than
millisecond pulsars in the rotation-powered state. Here, we model the
 X-ray and gamma-ray emission observed from {\1023} in such a
 state based on the assumption that most of the disk
in-flow is propelled away by the rapidly rotating neutron star
magnetosphere, and that electrons can be accelerated to energies of a
few GeV at the turbulent disk-magnetosphere boundary. We show that the
synchrotron and self-synchrotron Compton emission coming from such a
region, together with the hard disk emission typical of low states of
accreting compact objects, is able to explain the radiation observed
in the X-ray and gamma-ray band.  The average emission observed from
{\1023} is modelled by a disk in-flow with a rate of $1$--$3\times
10^{-11}$ M$_{\odot}$ yr$^{-1}$, truncated at a radius ranging between
30 and 45 km, compatible with the hypothesis of a propelling
magnetosphere. We compare the results we obtained with models that
rather assume that a rotation-powered pulsar is turned on, showing how
the spin down power released in similar scenarios is hardly able to
account for the magnitude of the observed emission.

\end{abstract}


\section{Introduction}

Millisecond pulsars are neutron stars (NS) that attained their quick
rotation during a Gyr-long phase of accretion of matter transferred
from a companion star, in a low-mass X-ray binary
\citep[LMXB;][]{bisnovatyikogan1974,alpar1982,radhakrishnan1982,wijnands1998}.
At the end of such X-ray bright phase, a pulsar powered by the
rotation of its magnetic field turns on, driving pulsed emission from
the radio to the gamma-ray energy band.
The recent discovery of {\m28}, a pulsar observed to swing between
accretion- (X-ray) and a rotation-powered (radio) pulsar regimes,
proved the tight evolutionary link existing between these two classes
of systems \citep{papitto2013nat}. {\m28} turned on in March
  2013 as an X-ray bright ($L_X\ga 10^{36}$ erg s$^{-1}$),
  accreting ms pulsar in the globular cluster M28; cross-referencing
  with catalogues of radio pulsars it was realised that the source
  behaved as a rotation powered radio pulsar a few years before. A few
  days after the end of the two-weeks long X-ray outburst, the source
  reactivated as a radio pulsar.
Variations of the mass in-flow rate towards the NS are able to drive
such state transitions.  During the rotation-powered regime, the
pressure of the pulsar ejects the matter transferred from the
companion star, causing irregular eclipses of the radio pulsed
emission.
Occasionally, the pressure of the matter
transferred from the companion overcomes the pulsar pressure, and
yields the formation of an accretion disk around the NS.

The state transitions observed from two more pulsars, {\1023}
\citep{archibald2009,patruno2014,stappers2014} and {\xss}
\citep{bassa2014,bogdanov2014,roy2014}, as well as in archival
observations of {\m28}
\citep{papitto2013nat,pallanca2013,linares2014}, showed that the
presence of an accretion disk does not necessarily imply the onset of
a bright X-ray outburst. 
These three transitional ms pulsars
showed an intermediate regime, which we dub {\it sub-luminous} disk
state, whose main features are 

\begin{itemize}
\item  the presence of an accretion disk,
as indicated by H$\alpha$ broad, sometimes double peaked emission
lines observed in the optical spectrum
\citep{wang2009,pallanca2013,halpern2013,demartino2014}; 

\item an average X-ray luminosity ranging from $10^{33}$ to $10^{34}$
  erg s$^{-1}$, intermediate between the level observed at the peak of
  X-ray outbursts ($10^{36}$ erg s$^{-1}$) and during the rotation
  powered emission ($<10^{32}$ erg s$^{-1}$); the X-ray emission is
  variable on time-scales of few tens of seconds and has a spectrum
  described by a power-law with index $\Gamma\simeq1.5$ and no cut-off
  below 100 keV
  \citep{saitou2009,demartino2010,demartino2013,papitto2013nat,linares2014,patruno2014,tendulkar2014}. {\m28}
  is the only transitional ms pulsar that has been observed
  also into an X-ray bright ($L_X\ga 10^{36}$ erg s$^{-1}$) outburst,
  so far.

\item presence of accretion-driven X-ray coherent pulsations at
  an rms amplitude between 5 and 10 per cent, that were detected from the two sources
  that were observed at a temporal resolution high enough during the {\it sub-luminous} disk state, {\1023}
  \citep{archibald2015} and {\xss} \citep{papitto2015}. A search for
  radio pulsations was conducted in the case of {\1023}, but none was
  detected with an upper limit on the pulsed flux one order of
  magnitude lower than during the rotation powered state
  \citep{stappers2014,bogdanov2015}; 

\item correlated variability of the X-ray and UV emission on
timescales of hundreds of seconds \citep{demartino2013}; 

\item a 0.1-10 GeV gamma-ray luminosity of $\approx10^{34}$ erg
  s$^{-1}$ detected from the two {\it transitional} millisecond
    pulsars of the Galactic field, {\1023} and {\xss}\footnote{Gamma-ray emission from the
      globular cluster hosting {\m28}, M28, has been detected by
      Fermi/LAT, but source confusion does not allow to pinpoint its
      origin in the cluster.}, up to ten times brighter with respect
  to the level observed during the rotation powered state
  (\citealt{demartino2010,hill2011}, Papitto, Torres \& Li 2014,
  \citealt{stappers2014,takata2014}). Candidate transitional ms
    pulsars have been also recently proposed to explain the otherwise
    unidentified gamma-ray sources 1FGL J0523.5-2529
    \citep{strader2014}, and 3FGL J1544.6-1125 \citep{bogdanov2015b}. 
           Transitional pulsars are the only low-mass X-ray
          binaries from which a gamma-ray emission has been detected
          so far by Fermi/LAT (see Sec.~\ref{sec:accrns} for a
            discussion of possible selection effects).

\item a bright, flat-spectrum radio emission indicative of
  partially absorbed synchrotron emission; transitional ms pulsars in
  this state are 1-2 orders of magnitude brighter at radio frequencies
  with respect to the extrapolation of the radio/X-ray correlation
  observed from X-ray brighter NS \citep{deller2014}.

\end{itemize}

X-ray flaring around a luminosity of about $10^{33}$--$10^{34}$ erg
s$^{-1}$, lasting from a few days to a few months, has been reported
also for a number of LMXB hosting a NS (Aql X-1,
\citealt{campana2003,cackett2011,cotizelati2014}; Cen X-4,
\citealt{rutledge2001,bernardini2013}; EXO 1745-248, \citealt{degenaar2012}; SAX
J1750.8-2900, \citealt{wijnands2013}; XMM J174457-2850.3,
\citealt{degenaar2014}). 
These observations suggest that a state intermediate between outburst
and quiescence is realized also in sources that did not show signs of
a magnetosphere so far.\footnote{Aql X-1 showed accretion-powered
  coherent pulsations only during a brief 150 s interval,
  \citealt{casella2008}.}  However, the presence of a spectral cut-off
at about 10 keV in the spectrum of Cen X-4 \citep{chakrabarty2014} and
the lack of a detection at gamma-ray energies indicate that in some of
these systems different radiation processes might be at work with
respect to  transitional ms pulsars in the {\it sub-luminous}
disk state.

The observation of coherent X-ray pulsations from {\1023} and {\xss}
during the disk state is most easily explained in terms of channelling
of at least part of the disk mass in-flow onto the magnetic poles
\citep[see][and Sec.~\ref{sec:discussion}
  below]{archibald2015,papitto2015}. For this to happen the disk
should penetrate into the light cylinder of the pulsar, ruling out the
the possibility that a rotation powered pulsar is turned on. However,
the exact nature of the energy reservoir that powers the emission of
{\it transitional} ms pulsars in the {\it sub-luminous} disk state is
still uncertain, as well as the mechanism that accelerate charges to
relativistic energies and yield the observed gamma-ray emission.
Papitto, Torres \& Li (2014, P14 in the following, see also
\citealt{bednarek2009}), proposed that a ms pulsar that prevents most
of the mass in-flow from accreting onto the NS surface due to its
rapid rotation (i.e., the propeller effect, \citealt{illarionov1975}),
could accelerate electrons at the boundary between the disk and the
propelling magnetosphere. If so happens, P14 showed that the
synchrotron emission yield by the interaction of electrons with the
field lines at the magnetospheric boundary makes a significant
contribution to the X-ray observed emission, and that gamma-rays are
produced by Compton up-scattering of the synchrotron photons.  Here,
we apply this model to explain the X-ray/gamma-ray emission observed
from {\1023} after its state transition to the {\it sub-luminous} disk
state in June 2013 \citep{patruno2014,stappers2014}, taking
explicitely into account the detection of coherent pulsations in the
X-ray light curve. {\1023} is particularly suited for the study of the
{\it sub-luminous} disk state, as parameters such as distance, spin
period and magnetic dipole moments are known with a high accuracy.

\section{{\1023}}
\label{sec:1023}

{\1023} was discovered in 2007 as a 1.7 ms radio pulsar in a 4.8 hr
orbit around a 0.2 M$_{\odot}$ companion star
\citep{archibald2009}. Observation of double peaked emission lines in
its optical spectrum indicated that in 2001 it likely had an accretion
disk \citep{wang2009}, suggesting that a state transition must have
occurred between then and 2007. The upper limit on the X-ray
luminosity when a disk was present ($2.7\times10^{34}$ erg s$^{-1}$
for a distance of 1.37 kpc, Deller et al. 2012) led
\citealt{archibald2009} to assume that mass in-fall was halted before
reaching the NS surface by the propeller inhibition of accretion.

During the radio pulsar state {\1023} showed a spin down luminosity of
$L_{sd}=4.4\times10^{34}$ erg s$^{-1}$ \citep{archibald2013}. Using
the relation given by \citet{spitkovsky2006}, the magnetic dipole is
estimated as $\mu=0.79\times10^{26}\;(1+\sin^2\alpha)^{-1}$ G cm$^3$ ,
where $\alpha$ is the angle betwen the magnetic and spin axis.  In the
radio pulsar state, the X-ray emission is described by a power law
with index $\Gamma=1.17(8)$ and a 3--79 keV luminosity of
$7.4(4)\times10^{32}$ erg s$^{-1}$ (\citealt{tendulkar2014}; see also
\citealt{bogdanov2011}), and the gamma-ray emission by a log-parabola
with $\beta=2.49(3)$ and 0.1-100 GeV luminosity of
$1.2(2)\times10^{33}$ erg s$^{-1}$ \citep{nolan2012}. The efficiency
of spin-down luminosity conversion in the considered bands is of 1.7
and 2.7 per cent, respectively.

\begin{figure}
\includegraphics[angle=0,width=\columnwidth]{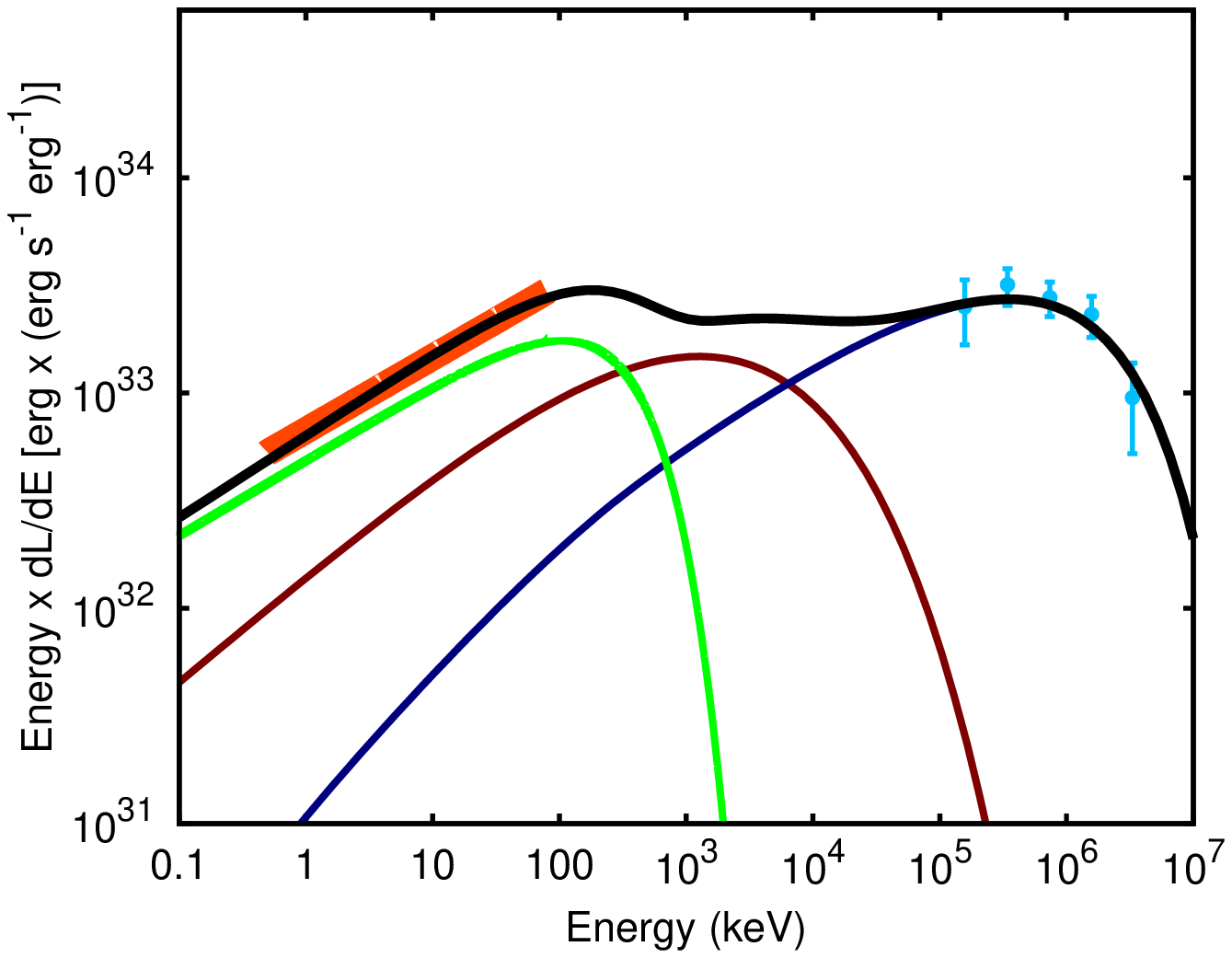}
\caption{Average SED observed from {\1023} in X-rays (orange strip,
  from \citealt{tendulkar2014}), and gamma-rays (cyan points,
  from \citealt{takata2014}), evaluated for a distance of $1.37$
  kpc. The total SED evaluated with our modelling for $\xi=0.15$
  (i.e., 15 per cent of the advected disk energy available to power
  the propeller emission), $\beta=0$ (anelastic propeller collision),
  $k_{ej}=0.99$ (i.e., 99 per cent of disk mass ejected), and
  $\omega_*=2.5$ is plotted as a black solid line. Synchrotron, SSC,
  and accretion flow (i.e. the sum of disk and NS emission) components are plotted as red, blue and green lines,
  respectively.
 \label{fig:sed}}
\end{figure}

In June 2013 the disaparance of radio pulsations at all orbital phases
\citep{stappers2014}, and the appearance of double peaked H$\alpha$
emission lines in the optical spectrum \citep{halpern2013} marked a
transition to an accretion, X-ray {\it sub-luminous} disk state. The
X-ray luminosity increased by one order of magnitude with respect to
the level shown in the radio pulsar state \citep{patruno2014}.
\citet{tendulkar2014} modelled with a power law with index
$\Gamma=1.66(6)$ the spectra observed in the disk state by the Swift
X-ray Telescope (XRT) and by NuStar in the 0.3-10 keV and 3-79 keV
energy bands, respectively (the orange strip in
Fig.~\ref{fig:sed}). The 3--79 keV luminosity was estimated as
$5.8(2)\times10^{33}$ erg s$^{-1}$, which corresponds to a luminosity
of $7.3\times10^{33}$ erg s$^{-1}$ in the 0.3--79 keV band, i.e. 16.6
per cent of the spin down power. 

During the {\it sub-luminous} disk state, three flux modes are
observed both in soft \citep{patruno2014,bogdanov2015} and in hard
X-rays \citep{tendulkar2014}. On the one hand, a low and a high state
characterized by a flux change of one order of magnitude. On the other
hand, the source occasionally presents an even brighter flaring
state. Coherent pulsations were detected at an rms amplitude of
$\simeq6\%$ when the source is in the high state, in which it spends
70\% of the time \citep{archibald2015}.  {\1023} brightened also in
gamma-rays, attaining a 0.1--100 GeV luminosity of
$(9.6\pm1.3)\times10^{33}$ erg s$^{-1}$, i.e. 22 per cent of the spin
down power \citep{takata2014}. The avergage Fermi LAT spectrum is
described by a power law with index $1.8(2)$, cut-off at an energy of
2.3(9) GeV (cyan points in Fig.~\ref{fig:sed}, taken from Fig.~2 of
\citealt{takata2014}).

\section{The model}

According to the standard theory of disk accretion onto a magnetized
rotator three regimes are realized depending on the radius $R_{in}$ at
which the disk is truncated by the rotating magnetosphere (see
\citealt{lipunov1987} for a review).
If $R_{in}$ is larger than the light-cylinder radius,
\begin{equation}
R_{lc}=cP/2\pi, 
\end{equation}
where $P$ is the spin period of the NS, the pulsar
emits a relativistic wind of particles and magneto-dipole radiation
that overcomes the gravitational force on the plasma transferred from
the companion star (the {\it ejector} state).
As the pulsar pressure declines with the distance from the
NS less steeply than the disk pressure, the emission of a pulsar in
such a state is expected to disrupt the entire disk flow
\citep{shvartsman1970,burderi2001}, even if stable solutions implying
the survival of the disk for $R_{in}\sim\mbox{few}\,R_{lc}$ have been
shown by \citet{eksi2005b}.

At a larger mass in-flow rate the in-falling plasma manages to
penetrate into the light-cylinder, and suppresses the relativistic
wind of particles and the rotation-powered pulsed emission. Accretion
onto the NS surface can take place freely only if the plasma in the
disk rotates faster than the NS magnetic field at the inner disk
boundaries, i.e. if $R_{in}$ is smaller than the co-rotation radius,
\begin{equation}
R_c=(GMP^2/4\pi^2)^{1/3}, 
\end{equation}
where $M$ is the NS mass, and Keplerian
rotation is assumed for disk plasma. Otherwise, if
$R_{lc}>R_{in}>R_{c}$, the accretion flow is repulsed by the quickly
rotating magnetosphere and can be ejected from the system, i.e. the
system is assumed to be in the propeller state \citep{illarionov1975}.

A useful parameterization of the problem is obtained introducing
  the dimensionless fastness,
\begin{equation}
\label{eq:omega}
\omega_* \equiv \frac{\Omega_*}{\Omega_K(R_{in})}= \left( \frac{R_{in}}{R_c} \right)^{3/2}
\end{equation}
\citep{ghosh1977} where $\Omega_K$ is the Keplerian angular frequency
and $\Omega_*$ is the NS angular frequency. For values of $\omega_*$
ranging between 1 and 1.26, not enough kinetic energy is given to
plasma to eject it from the system; matter is then expected to
build-up in the disk, and eventually overcome the pressure barrier set
by the magnetosphere, leading to a burst of accretion
\citep{spruit1993,dangelo2010}. Larger values of $\omega_*$ ranging
between 1.3 and 2.5 have been considered by \citet{lii2014} to perform
axisymmetric magneto-hydrodynamical simulations of accretion onto a
rapidly rotating magnetized star, driven by magneto-rotational
instability. While they also observed cycles between matter
accumulation at the inner disk boundary and episodes of accretion onto
the star, they  found that part of the inflow was ejected by the
quickly rotating magnetosphere. The fraction of the matter actually
ejected by the system with respect to that accreted was found to
increase continuously with $\omega_*$.

The determination of an analytical relation for $R_{in}$ as a
function of the main physical quantities characterizing such systems
(such as the disk mass in-flow rate $\dot{M}_{d}$, the magnetic moment
$\mu$, the mass $M_*$, and the spin period $P$ of the magnetized
rotator), and of the physics of the field-disk interaction, is one of
the open problems of the current theoretical investigation in the
field \citep[see, e.g.,][]{ghosh1979,wang1996,bozzo2009}. Usually the inner
disk radius is expressed as a fraction $k_{m}$ of the Alfven radius
$R_{A}$, a scale size obtained by equating the in-ward pressure of an
assumed spherical inflow to the outward pressure of a dipolar
magnetosphere:
\begin{equation}
\label{eq:rin}
R_{in}=k_{m}R_{A}=k_{m}\left[\frac{\mu^4}{2GM_*\dot{M}_d^2} \right]^{1/7}.
\end{equation}
Values of $k_{m}$ ranging from 0.5 to 1 have been proposed by a number
of authors depending on the details of the physics of the
disk-magnetosphere interactions \citep[see][and references
  therein]{ghosh2007}. \citet{lii2014} found in 3D MHD simulations of
disk accretion around a fast rotator that a value of $k_m=0.7$ 
matched the disc truncation radius reproduced by their modelling.  

\label{sec:xraypulse}
Assuming that the coherent X-ray pulsations observed from {\1023}
  in the disk state were due to accretion of matter onto a fraction of
  the NS surface, we can estimate an upper and a lower limit on the
  mass accretion rate onto the NS. The latter, $\dot{M}_{NS}$, is
  constrained assuming either that the whole observed X-ray luminosity
  is due to accretion onto the NS, or just that the pulsed
  luminosity is. Considering that $L_X(0.3-79$ keV)=7.3$\times10^{33}$
  erg s$^{-1}$ we then obtain
\begin{eqnarray}
\label{eq:mdotlimits}
5\times10^{-14}\mbox{M}_{\odot}\, \mbox{yr}^{-1}  & \simeq  & (\sqrt{2}A_{rms})\frac{L_X R_{NS}}{GM_{NS}} \nonumber \\
< & \dot{M}_{NS} & <  \nonumber \\
\frac{L_X R_{NS}}{GM_{NS}}  & \simeq  & 6 \times10^{-13}
\mbox{M}_{\odot}\, \mbox{yr}^{-1},
\end{eqnarray}
where $A_{rms}\simeq0.06$ is the rms amplitude of X-ray pulsations
\citep{archibald2015}, and we assumed a NS mass of 1.4 $M_{\odot}$ and
a radius of 10 km, as throughout the paper, and an efficiency of
conversion of accretion power into observed X-rays of unity.  Given a
magnetic moment of $\mu=0.79\times10^{26}$ G cm$^{3}$ \citep[see][and
  Sec.~\ref{sec:1023}]{archibald2013} and $k_m=0.7$ \citep{lii2014}, in order to keep
the inner disk radius within a few times the corotation radius (23.8
km for {\1023}, \citealt{archibald2009}), Eq.~\ref{eq:rin}
indicates that the disk mass accretion rate must be
\begin{equation}
\dot{M}_d\simeq 7\times10^{-11}\,\mbox{M}_{\odot}\,
\mbox{yr}^{-1}>100\times \dot M_{NS}. 
\end{equation}
By interpreting X-ray pulsations as
due to accretion onto the NS magnetic poles, and assuming the the
inner disk radius can be reproduced by a relation like
Eq.~\ref{eq:rin}, it then follows that $\ga 99$ \% of the matter in-flowing
in the disk must be ejected from the inner disk boundary for mass
conservation to hold.

In P14 we developed a model to interpret the emission of millisecond
transitional pulsars in the sub-luminous disk state assuming that:

\begin{itemize}

\item {  accretion onto the NS surface is inhibited by the propeller effect
(i.e. $R_{lc}>R_{in}>R_{c}$); }

\item  { electrons are accelerated to
relativistic energies at
the turbulent boundary between the disk and the propelling
magnetosphere; }

\item  {relativistic electrons interact with the NS magnetic
  field lines producing synchrotron emission that explains (at least part of)
  the X-ray emission;}

\item  {synchrotron photons are up-scattered by relativistic
  electrons, to explain the emission observed in the gamma-ray band.}

\end{itemize}

Here, we apply a similar model to the case of {\1023}, restricting to
a range of values of fastness ranging from 1.5 to 2.5, in order for
the inner disk radius to be close enough to the corotation radius to
let a fraction of matter effectively accrete down to the NS surface, as observed.

\subsection{The energy budget}

Energy conservation dictates that the energy available to power the
radiative emission from the disk ($L_{disk}$), the inner disk boundary
($L_{prop}$) and the NS surface ($L_{NS}$), as well as the
kinetic energy of the outflow launched by a propelling NS
($\dot{M}_{ej}v_{out}^2/2$) and the energy converted in internal
energy of the flow and advected ($\dot{E}_{adv}$), should be equal to
the sum of the  gravitational energy liberated by the in-fall of
  matter $\dot{E}_g$, and the energy released by the NS magnetosphere
through the torque $N$ applied at the inner disk radius:
\begin{equation}
\label{eq:encons}
L_{prop}+L_{d}+\dot{E}_{adv}+L_{NS}+\frac{1}{2}\dot{M}_{ej}v_{out}^2=\dot{E}_g+N \Omega_*,
\end{equation}
where $\Omega_*=2\pi/P$ is, as before, the NS angular velocity.  In a
  stationary state, mass conservation is expressed by:
\begin{equation}
\dot{M}_d=\dot{M}_{NS}+\dot M_{ej}.
\end{equation}
We define the fraction of mass ejected as
\begin{equation}
k_{ej}=\frac{\dot{M_{ej}}}{\dot{M}_d}=1-\frac{\dot{M}_{NS}}{\dot{M}_d}.
\end{equation}
Considering the reasoning developed in the previous section, we
consider values $k_{ej}>0.95$ for the case of {\1023}.  
According to
these definitions, the gravitational energy liberated by the mass
in-fall is:
\begin{equation}
\label{eq:eg}
\dot{E}_g=\frac{GM\dot{M}_d}{R_{in}}+{GM\dot{M}_{NS}}\left(\frac{1}{R_{NS}}-\frac{1}{R_{in}}\right).
\end{equation}

We assume that the NS luminosity is given by efficient conversion of
the in-falling gravitational energy, so that
\begin{equation}
\label{eq:lns}
L_{NS}=GM\dot{M}_{NS} \left( \frac {1}{R_{NS}}-\frac{1}{R_{in} } \right).
\end{equation} Furthermore, we
express the disk luminosity as a fraction $\eta$ of the energy radiated
by an optically thick, geometrically thin disk:
\begin{equation}
\label{eq:ldisk}
L_{d}=\eta\frac{GM\dot{M}_d}{2R_{in}}.
\end{equation}
The case of a radiatively efficient disk is realized for
$\eta=1$.  For values of $\eta$ lower than unity, we assume that
  the energy that is not radiated by the disk is partly advected, and
  partly made available to power the propeller emission. To express
  the latter, we introduce a parameter $\xi$, that will represent the
  fraction of gravitational energy liberated in the disk that can be
  used to power the propeller emission; for $\xi=0$ no such energy is
  used, while for $\xi=1-\eta$, all the disk energy that is not
  radiated is converted into propeller luminosity.  This implicitely
means that we express
  \begin{equation}
  \dot{E}_{adv} = (1- \eta - \xi) \frac{GM\dot{M}_d}{2R_{in}}.
  \end{equation}
  Substituting the previous formulae in Eq.~\ref{eq:encons}, we obtain an expression for the
energy that can be radiated from the disk-magnetospheric boundary:
\begin{equation}
L_{prop}=\left(\frac{1+\xi}{2}\right) \frac{GM\dot{M}_d}{R_{in}}+N\Omega_* -\frac{1}{2}k_{ej}\dot{M}_{d}v_{out}^2.
\label{eq:encons2}
\end{equation}

\begin{figure}
\includegraphics[angle=0,width=\columnwidth]{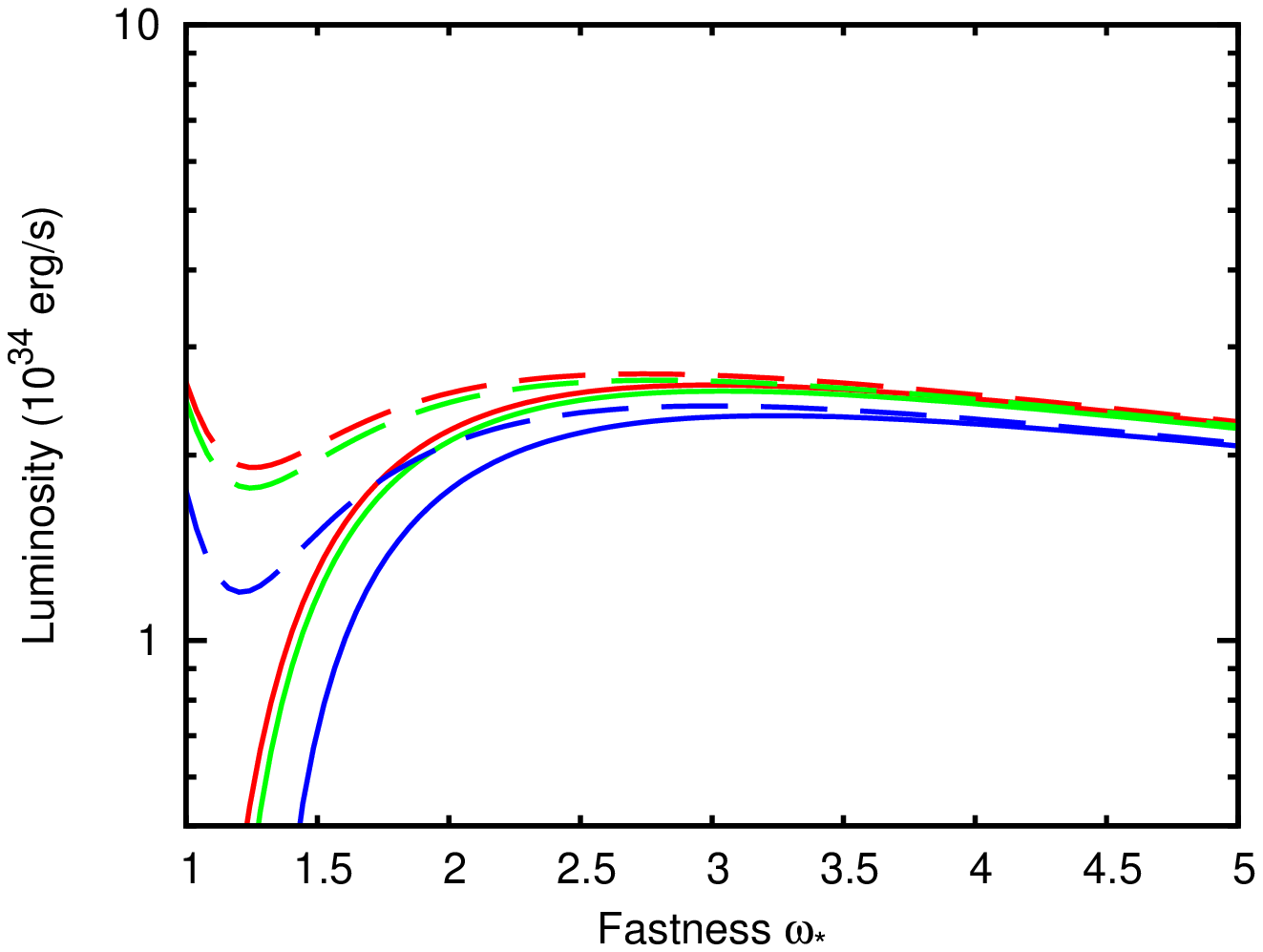}
\caption{\label{fig:lum} Propeller luminosity in the case of almost
  total ejection ($k_{ej}=1$), and $\xi=0$ (i.e., no disk
  gravitational energy available to power the propeller emission, red
  solid line), and 0.15 (red dashed line). The green solid and
  dashed lines refer to the case $k_{ej}=0.99$ evaluated for $\xi=0$
  and 0.15, respectively. The blue solid and dashed lines refer to the
  case $k_{ej}=0.95$ evaluated for $\xi=0$ and 0.15, respectively.
  All the curves are plotted for a totally anelastic interaction
  ($\beta=0$).}
\end{figure}

Similarly, the conservation of angular momentum at the inner disk boundary
yields:
\begin{equation}
\dot{M}_{ej}R_{in}v_{out}=N+\dot{M}_{d}\Omega_K R_{in}^2.
\label{eq:amcons}
\end{equation}
The left hand side represents the angular momentum imparted to
  eject matter, while on the right hand side the torque exerted by the
  NS magnetosphere and the angular momentum advected by disk matter in
  Keplerian rotation appear. In order to express the velocity of the
outflow $v_{out}$ we follow \citet{eksi2005}, who treated the
interaction at the inner disk boundary as a collision of
particles,obtaining:
\begin{equation}
\label{vout}
v_{out}=\Omega_K(R_{in}) R_{in}[\omega_*(1+\beta)-\beta],
\end{equation}
where the elasticity parameter $\beta$ has been introduced. The case
of completely anelastic collision is given by $\beta=0$, while the
totally elastic case is described by $\beta=1$.  Inserting the
expression for the outflow velocity (Eq.~\ref{vout}) into
Eqs.~\ref{eq:encons2} and \ref{eq:amcons}, and solving for $L_{prop}$
and $N$ gives:
\begin{eqnarray}
\label{eq:sys}
L_{prop}&=&\frac{GM\dot{M}_d}{R_{\rm in}} 
\left\{ \frac{1+\xi}{2} -\omega_* + k_{ej}
\right.
\nonumber \\ && 
\left.
\left[\omega_*[\omega_*(1+\beta)-\beta ]-\frac{1}{2}[\omega_*(1+\beta)-\beta]^2 \right] \right\} \nonumber \\
N&=&\dot{M}_d\sqrt{G M R_{in}} \{ k_{ej}[\omega_*(1+\beta)-\beta]-1\}. \label{eq:torque}
\end{eqnarray}
If most of the in-flowing matter is ejected from the inner disk
boundary ($k_{ej}\simeq1$), the equation for the propeller luminosity simplifies to:
\begin{equation}
L_{prop}=\frac{GM\dot{M_d}}{2R_{\rm in}} [\xi+(\omega_*-1)^2(1-\beta^2)].
\end{equation}  It is immediate
to note that if the energy advected in the disk is not available to
power the propeller ($\xi=0$) and the propeller collision is
completely elastic ($\beta=1$), all the available energy goes into
powering the kinetic energy of the outflow, and the propeller
luminosity vanishes. On the other hand, if the collision is completely
anelastic ($\beta=0$), the outflow velocity equals the velocity of the
magnetosphere at the inner disk radius, and the propeller luminosity
is maximized. In the following we only consider the latter case. Using
Eq.\ref{eq:rin} to relate the disk mass accretion rate to the inner
disk radius, and Eq.~\ref{eq:omega} to express the latter in terms of
the corotation radius and the fastness, we obtain for the parameters
of {\1023}:
\begin{equation}
L_{prop}=1.75\times10^{35}\,\omega_*^{-3}[\xi+(\omega_*-1)^2(1-\beta^2)]\,\mbox{erg s}^{-1}.
\end{equation}
This relation is plotted in Fig.~\ref{fig:lum} for $\beta=0$ and $\xi$
equal to 0 (i.e., no disk gravitational energy available to power the
propeller emission, red solid line), and 0.15 (red dashed line). Green
and blue lines in the same figure represent the cases $k_{ej}=0.99$
and 0.95, respectively.  Values of the propeller luminosity of few
$\times 10^{34}$ erg s$^{-1}$ are obtained for a fastness $\ga 1.5$.

\subsection{The electron population}

Like in P14, we assume that at the inner disk boundary the propelling
magnetosphere is able to accelerate electrons to relativistic
energies. To simplify, the electron energy distribution is assumed to
be described by a power-law with an exponential cutoff:
\begin{equation}
\frac{dN_e}{d\gamma}=N_0 \gamma^{-\alpha}
\exp{\left(-\frac{\gamma}{\gamma_{max}}\right)}.
\end{equation}
These electrons produce synchrotron emission by interacting with the
magnetic field at the disk-magnetosphere interface:
\begin{equation}
\label{eq:B}
\bar{B}=\frac{\mu}{R_{in}^3}=\frac{\mu}{R_c^3 \omega_*^2}.
\end{equation}
At high-energies the synchrotron emission is cut-off at $E_{syn}\simeq
10 (\bar{B}/10^6\, \mbox{G})\,(\gamma_{max}/10^4)^2$ MeV, while at low
energy is self-absorbed below a few eVs (see \citealt{papitto2014} and
references therein). We further assume that the electron distribution
lies in a region of volume $V$ (giving an electron density
$n_e=N_0/V$), and up-scatters the synchrotron photons up to an energy
$E_{ssc}=\gamma_{max}m_ec^2\simeq5 (\gamma_{max}/10^4)$ GeV, to give a
self-synchrotron Compton (SSC) component. Therefore, in our model, the
X-ray emission is given by the sum of two components; the synchrotron
emission powered by the propeller luminosity $L_{prop}$, and the
  X-rays produced by the accretion flow, either from the disk, the NS
  surface or a corona surrounding the system.  On the other hand, the
gamma-ray emission falling in the 0.1-10 GeV range could only arise as
the SSC component of the propeller emission.

The parameters of the electron distribution ($\alpha$, $\gamma_{max}$,
$n_e$) and the volume $V$ of the region of acceleration can be then
adjusted such that: \\
(i) they model the gamma-ray emission as the
self-synchrotron Compton component of the same electron population
that produces a good part of the X-ray emission; and \\
(ii) for a fixed
value of the fastness $\omega_*$, they give a total propeller luminosity (i.e, the
sum of synchrotron and self-synchrotron Compton contributions) that
matches the value given above by Eq.~\ref{eq:sys}.

As the X-ray emission is given by the sum of the contribution of
  the propeller and the accretion flow luminosity in that energy
  range, and no cut-off is observed from 0.5 to 80 keV, we cannot
  disentangle the relative weight of these components.  At low
  luminosities ($\la 10^{34}$--$10^{35}$ erg s$^{-1}$), the X-ray
  spectrum of the accretion flow onto compact objects is generally
  dominated by a power law with index $\Gamma$ between 1.5 and 2.5,
  extending to high energies ($\ga 100$ keV; see, e.g.,
  \citealt{reynolds2010}, \citealt{armaspadilla2013a}a, \citealt{barmaspadilla}b). A
  similar spectral shape is also observed for the pulsed emission of
  accreting millisecond pulsars \citep[see][and references
    therein]{patruno2012c}.  We then describe the contribution of the
accretion flow in the X-ray band as a power-law with index
  $\Gamma=1.65$, cut-off at an energy outside the observed energy
band (we arbitrarily chose 300 keV as an example), that summed to the
synchrotron and SSC component gives an 0.3-79 keV X-ray flux equal to
the one observed in that energy band. Then, we require that once
  the propeller contribution to the X-ray emission is fixed by our
  modelling of the gamma-rays, the accretion flow X-ray emission does not
  exceed the observed emission.  A lower limit on the contribution of
  the accretion flow to the X-ray flux is given by the  pulsed flux
  amplitude, $\sqrt{2}A_{rms}L_X\simeq 6\times10^{32}$ erg s$^{-1}$,
  while an upper limit to the accretion flow X-ray luminosity is
  given by the total luminosity observed in that band ($7.3\times10^{33}$
  erg s$^{-1}$).

\section{Results}


To simulate the emission processes that take place in a magnetized
environment filled by a relativistic population of electrons we used
the codes described by \citet{torres2004,decea2009,martin2012}.  The
cut-off observed at few GeVs in the Fermi-LAT energy band, requires a
maximum electron energy of $\gamma_{max}=10^4$. We also set the
  index of the electron distribution as $\alpha=2$.

 In order to have enough energy to power the propeller emission
 necessary to explain the observed gamma-ray emission, we fixed the
 elasticty parameter to 0 (i.e., inelastic scattering), allowed for 15
 per cent of the energy advected in the disk to be used to power the
 propeller emission (i.e. $\xi=0.15$), and set $k_{ej}=0.99$
 (i.e. only 1 per cent of the disk mass is accreted onto the NS, while
 the rest is ejected). This gives a maximum propeller luminosity of
 $\ga 2.5\times10^{34}$ erg s$^{-1}$ for value of the fastness between
 1.5 and 2.5 (see Fig. \ref{fig:lum}).  We modelled the spectral
 energy distribution considering different values of the fastness in
 this range.
   
   As the co-rotation radius and the dipole magnetic moment of {\1023}
   are measured; the inner disk radius, the strength of the magnetic
   field at the disk inner boundary, and the disk mass accretion rate
   are fixed by the choice of $\omega_*$ (see Eqs.~\ref{eq:omega},
   \ref{eq:B} and \ref{eq:rin}, respectively). For a given value of
   $\omega_*$, we then evaluate the electron density $n_e$ and the
   volume $V$ of the acceleration region requested to describe the
   observed gamma-ray spectrum with the SSC component (which requires
   a luminosity of $L_{SSC}\simeq 1.7\times10^{34}$ erg s$^{-1}$), and
   to give a total propeller luminosity equal to the value given by
   Eq.~\ref{eq:sys} when $L_{SSC}$ is summed to the synchrotron
   component. The values of the parameters obtained for different
   values of $\omega_*$ are listed in Table 1, while the spectral
   energy distribution obtained for $\omega_*=2.5$ is plotted in
   Fig.~2, as an example. The X-ray luminosity attributed to the
   accretion flow, $L^X_{accr}$, is evaluated as the difference
   between the luminosity observed in the 0.3-79 keV band
   ($7.3\times10^{33}$ erg s$^{-1}$) and the sum of the synchrotron
   and SSC luminosity falling in the same energy band.
 
For values of $\omega_*$ close to 1.5, the propeller luminosity is
lower, and in order to explain the observed spectral energy
distribution with our modelling, a large electron density and a very
small volume of emission of the SSC component are required,
corresponding to a typical size of the emission region of $\approx
0.05$ km. Increasing $\omega_*$ a larger region with typical size of
$\la 1$ km is instead allowed. We discuss more on this below.

We checked that the assumption that only 1 per cent of the disk mass
is accreted is compatible with the parameters of our modelling, by
evaluating the ratio between the NS mass accretion rate (constrained
by using Eq.~\ref{eq:mdotlimits}), and the values of the disk mass
accretion rate listed in Table 1.  For all the models listed in Table
1 this yielded values of $k_{ej}$ ranging between 0.94 and 0.998, thus
compatible with the assumed value of 0.99.

The efficiency of conversion of gravitational energy into X-rays was
estimated as $\eta_X=L^X_{accr}/\dot{E}_g$. Plugging the parameters of
our models into Eq.~\ref{eq:eg}, we obtain values of the efficiency
ranging between $\approx 5$ and $20$ per cent, with larger efficiency
values obtained for the models with the larger fastness. As the X-ray
efficiency owing to the in-fall of matter onto the NS surface is
unlikely much lower than one, we conclude that the accretion disk
efficiency must be lower than 20 per cent (note that according to our
definitions, theefficiency of a geometrically thin, optically thick
disk that emits all of is energy in X-rays is $\eta=0.5$).

\begin{deluxetable*}{cccccccccccc}
\tablecaption{Model parameters used to model the SED of {\1023} and
    {\xss}.  \label{tab}}

\tablehead{$\xi$ & $k_{ej}$ & $\omega_*$ & $R_{in}$ (km) & $\bar{B}$ (MG) &  $\dot{M}$   & $L_{prop}$ & $n_e$ ($10^{18}$ cm$^{-3}$) & V ($10^{15}$ cm$^3$) & $L_{ssc}/L_{sync}$ & $L^X_{accr}$ & $\eta^{X}_{accr}$}
\startdata
\cutinhead{{\1023}}

0.15 & 0.99 & 1.50 & 31.2 & 2.6 & 2.7 & 1.96 & 54 & $6\times10^{-4}$ & 5.0 & 0.65 & 0.06 \\
0.15 & 0.99 & 1.75 & 34.6 & 1.9 & 1.9 & 2.23 & 10 & 0.01 & 2.8 & 0.59 & 0.08 \\
0.15 & 0.99 & 2.00 & 37.8 & 1.5 & 1.4 & 2.43 & 5.0 & 0.04 & 2.4 & 0.55 & 0.11 \\
0.15 & 0.99 & 2.25 & 40.9 & 1.15 & 1.1 & 2.56 & 2.1 & 0.19 & 2.04 & 0.51 & 0.14 \\
0.15 & 0.99 & 2.50 & 43.8 & 0.94 & 0.8 & 2.62 & 1.3 & 0.50 & 1.9 & 0.48 & 0.17 \\
\cutinhead{{\xss}}
0.15 & 0.99 & 2.50 & 43.8 & 1.34 & 2.4 & 2.62 & 1.7 & 0.21 & 2.2 & 0.55 & 0.08 \
\enddata
\tablecomments{Input parameters are listed in the leftmost three columns. Physical quantities obtained using the analytical relations given in text, are listed in columns 4-8. Parameters estimated from the modelling of the observed SED are given in the {five} rightmost columns. Luminosities are given in units of $10^{34}$ erg s$^{-1}$, while the mass in-flow rate is expressed in units of $10^{-11}$ M$_{\odot}$ yr$^{-1}$.}
\end{deluxetable*}

\subsection{A comparison with {\xss}}

\begin{figure}
\includegraphics[angle=0,width=\columnwidth]{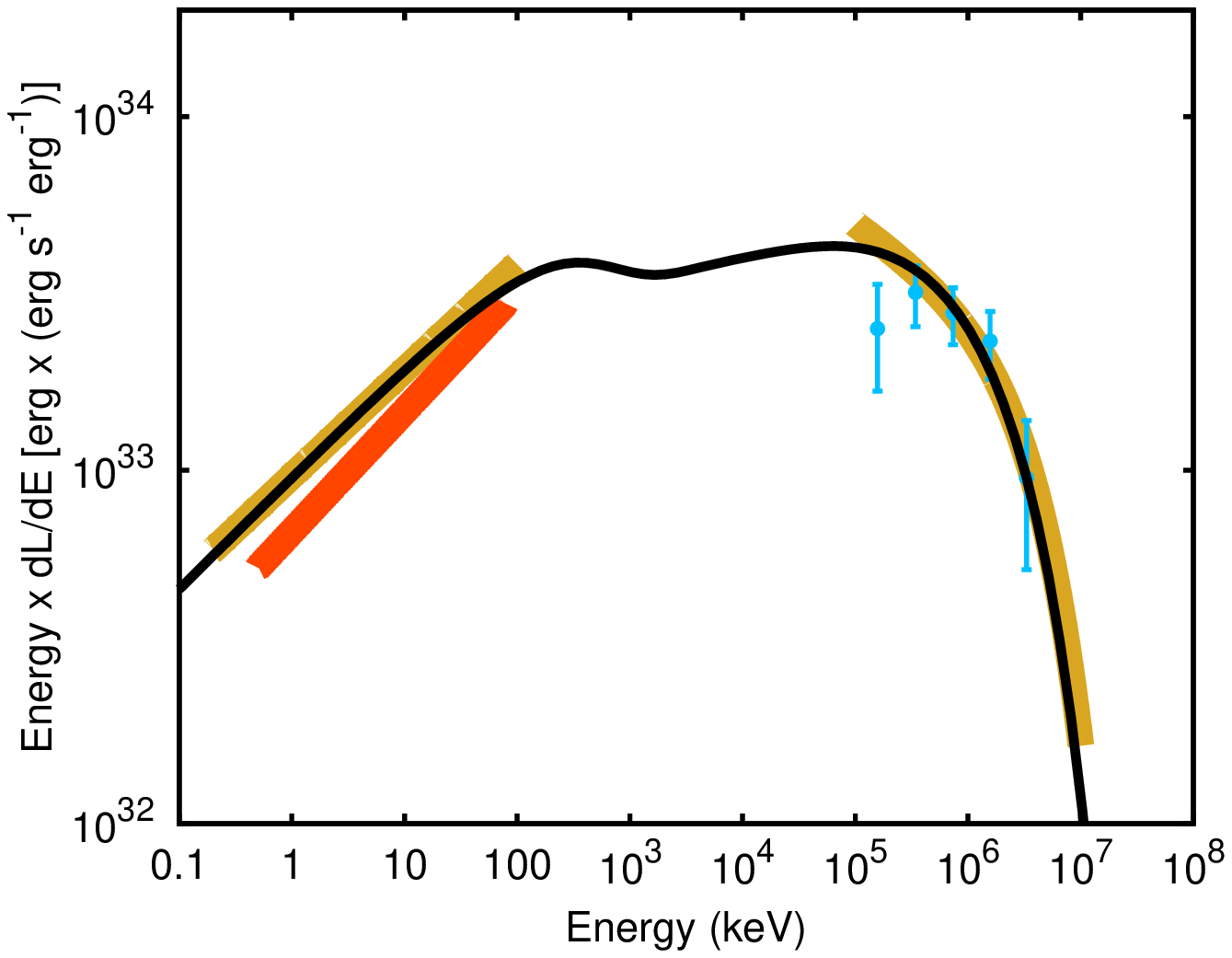}
\caption{ Average SED observed from {\1023} (orange stripe and cyan
  points) and {\xss} (yellow stripe; see \citealt{papitto2014} and
  references therein) for a distance of 1.4 kpc. The model of the
  {\xss} SED, obtained for $\xi=0.15$, $k_{ej}=0.99$ and
  $\omega_*=2.5$ (see Table~\ref{tab}) is plotted as a black solid line.
 \label{fig:compsed}}
\end{figure}

The generic idea behind the model used here to interpret the emission
observed from {\1023} in the sub-luminous disk state has been
previously successfully applied to the emission of {\xss} in the same
state (P14). However, in that paper it was assumed that all the
gravitational energy gave off by the matter in-flow was available to
power the propeller emission and outflow, and consequently, the
interplay between the disk and the propeller luminosities was not
explored as done above.  Considering the formalism developed in
Sec. 3, that assumption corresponds to the case $\eta=0$ and
$\xi=1$. This case is extreme and probably unlikely, since it would
require an X-ray {\it dark} accretion flow.

However, the similarities between the average SED observed from these
two sources, and the results obtained in the case of {\1023} (see
Sec. 4), indicate that a our modelling works also in the case of
{\xss}.  This is emphasized in Fig.~\ref{fig:compsed}, where the high
energy SED of {\xss} is overplotted as a yellow stripe to the SED of
{\1023}. A distance of 1.4 kpc was considered, equal to the value
indicated by the dispersion measure of the radio pulsed signal
observed by \citet{roy2014}, and at the lower bound of the 1.4-3.6 kpc
range determined by \citet{demartino2013}.

The spin period of {\xss} is very similar to that of {\1023} ($P=1.69$
ms), and the magnetic moment can be estimated as
  $\mu_{26}=1.13\,I_{45}^{1/2}\,(1+\sin^2\alpha)^{-1/2}$ G cm$^3$ from the
  observed spin down rate \citep{roy2014}.  We modelled the SED of
  {\xss} using similar parameters than those used for {\1023}, and
  $\omega_*=2.5$ (see Table~1), obtaining the model plotted as a
black solid line in Fig.~\ref{fig:compsed}.

\section{  { Discussion} }
\label{sec:discussion}

From a theoretical point of view, three possibilities are open to
explain the complex phenomenology of {\1023} (and possibly other
transitional pulsars) in the sub-luminous disk state: a
rotation-powered pulsar, accretion onto the NS surface, and a
propeller regime. In the following we discuss these possibilities on
the basis of the phenomenology observed.

\subsection{ Rotation-powered pulsar scenario} 

\citet{stappers2014}, \citet{takata2014}, \citet{cotizelati2014} and
\citet{Li2014} discussed the phenomenology observed from {\1023} in
terms of a rotation-powered pulsar active even in the presence of an
accretion disk, the radio coherent pulsation being washed out by the
enshrouding of the system by intra-binary material. According to these
models, the relativistic wind of particles emitted by the pulsar in
this state is responsible for the emitted gamma-rays.

\citet{stappers2014} and \citet{cotizelati2014} proposed
that the shock between the pulsar wind of particles and the mass
in-flow would be the most likely region where the gamma-ray emission
is generated, as this would explain the correlation between the
observed increase of the gamma-ray emission and the formation of a
disk in the system. On the other hand, \citet{takata2014} and
\citet{Li2014} interpreted the gamma-ray emission observed in the
sub-luminous disk state through inverse Compton scattering of UV disk
photons off the cold pulsar wind. Furthermore they assumed that the
X-rays were due to synchrotron emission taking place in the shock
between the pulsar wind and the plasma in-flow. Such a shock would be
expected to be stronger in the subluminous disk state than in the
pulsar state, as a larger fraction of the pulsar wind would be
intercepted, thus yielding the increased X-ray emission observed.

These ideas are similar to those used for gamma-ray binaries such as
LS 5039 or LS I +61 303, where also wind and inter-wind shocks are
studied as possible providers of accelerated electrons (e.g.,
\citealt{dubus2006,sierpowska2008}).

The main problematic aspect in applying any rotation-powered scenario
to the sub-luminous disk state of {\1023} is the observation of
  coherent X-ray pulsations with an rms amplitude of $\approx 6$ per
  cent during the accretion disk state. Even if X-ray pulsations were
  observed also during the rotation-powered state, the pulsed 0.5-10
  keV X-ray luminosity was $1.2\times10^{31}$ erg s$^{-1}$
  \citep{archibald2010}, roughly one order of magnitude lower than the
  pulsed luminosity observed during the disk state, $2.5\times10^{32}$
  erg s$^{-1}$ \citep{archibald2015}. Synchronous switching of the
  radio and X-ray pulsations properties has been observed from
  rotation-powered pulsars, and interpreted as due to global changes
  of the magnetospheric conditions \citep{hermsen2013}. However, to
  interpret the X-ray pulsations observed from {\1023} in the disk
  state as rotation-powered, one should admit that when a disk is
  present the pulsed flux increases by more than one order of
  magnitude with respect to the case of an unperturbed
  magnetosphere. This is contrary to the expectation that high-density
  plasma from the accretion process shorts out the electric fields
  which power the electron/positron acceleration in the vacuum gaps,
  switching off rotation-powered pulsations \citep[see,
    e.g.][]{illarionov1975}. We consider such a scenario as highly
  unlikely. Furthermore, the X-ray pulsations observed from {\1023}
  during the disk state are sinusoidal and have a non-thermal energy
  distribution \citep{archibald2015}, whereas non-thermal X-ray pulsations
  observed from rotation-powered pulsars are typically narrow peaked
  \citep{zavlin2007}.

In addition to X-ray pulsation, the high conversion efficiency of the
spin down power required to explain the observed radiation also
  disfavors a rotation-powered scenario. If a radio-pulsar
is switched on, the spin-down power of $4.4\times10^{34}$ erg s$^{-1}$
is the dominant source of energy for the system; as a matter of fact,
for the radio-pulsar to be active the inner disk radius must lie
beyond the light-cylinder radius $R_{lc}$ (80.6 km in the case of
{\1023}), and in such a case, the implied mass in-flow rate of less
than $\approx 10^{-12}$ M$_{\odot}$ yr$^{-1}$ (see Eq.~\ref{eq:rin})
would yield an accretion luminosity of $<10^{33}$ erg s$^{-1}$. The
sum of the average luminosity observed from {\1023} in the 0.3--79 keV
and 0.1--100 GeV energy bands amounts to $\simeq 1.7\times10^{34}$ erg
s$^{-1}$, a value that implies a spin-down power conversion efficiency
of $>40\%$. A similar value is already larger than the the values
observed from rotation-powered pulsars, which typically convert 0.1
\citep{possenti2002,vink2011} and 10 per cent \citep{abdo2013} of their spin down
power into observable X-rays and gamma-rays, respectively. In
addition, if one considers that the spectral energy distribution is
most likely flat (or even peaks) in the 1-10 MeV energy range where
observations lacks (see Fig.~\ref{fig:sed} of this paper, and Fig.~18
of \citealt{tendulkar2014}), the total power obtained modelling the
SED with two smooth components easily attains a value of the order of
the spin down power or larger (indeed, the power of most of the models
listed in Table 1 exceeds such a threshold). Note that a total
luminosity significantly exceeding the spin down power would directly
exclude a rotation-powered scenario.

Furthermore, the strong flickering observed in X-rays makes the case
for the spin-down power being the lone source of energy even more
unlikely. The peak observed X-ray luminosity is of the order of
$\approx 2\times10^{34}$ erg s$^{-1}$
\citep{patruno2014,tendulkar2014}, i.e. roughly 40 per cent of the
spin down power, alone. No information is available about the
correlation between the X-ray and the gamma-ray luminosity on short
time-scales, but it is clear that unless they are strictly
anti-correlated, this likely implies the limit set by the spin down
power to be exceeded at the peak of flares.  Furthermore, the power
emitted through synchrotron emission only depends on the acceleration
efficiency, the solid angle under which the shock is seen by the
isotropic pulsar wind, and the bow-shock geometry
\citep{arons1993}. It seems highly unlikely that these may produce the
variability of the X-ray emission by up to two orders of magnitude
observed over time-scales of few tens of seconds from {\1023}.

\subsection{Accreting NS scenario} 
\label{sec:accrns}

The most immediate interpretation of the coherent X-ray
  pulsations observed from {\1023} is then in terms of accretion of at
  least part of the disk in-flow close to the NS magnetic poles.
  However, in Sec.~\ref{sec:xraypulse} we already showed that
  assuming that the observed average X-ray luminosity $L_X$ is
entirely due to accretion onto the NS surface, the implied mass
accretion rate onto the NS surface is $\dot{M}_{NS}<6\times10^{-13}$
M$_{\odot}$ yr$^{-1}$ (see Eq.~\ref{eq:mdotlimits}). According to
  Eq.~\ref{eq:rin}, and considering the value of the dipole magnetic
  moment of {\1023} measured from the spin down rate, at such a mass
  accretion rate the disk would be truncated beyond the light-cylinder
  radius. Such a large value clearly violates the criterion for
accretion onto the NS surface to proceed.  In order to keep the
  accretion disk radius closer to the co-rotation radius, we thus had
  to assume that the disk mass accretion rate was much larger than the
  rate at which mass is effectively accreted onto the NS, and that the
  excess disk mass is ejected by the system. 

In addition, the simultaneous observation of a bright gamma-ray
emission would be unexplained by a fully-accreting scenario,
considering that the transitonal pulsars {\1023} and {\xss} are the
only LMXB from which a Fermi gamma-ray counterpart could be securely
identified among a population of $> 200$ known accreting LMXB,
  and this is unlikely to be the result of a selection effect. The
  5-$\sigma$ sensitivity flux level above 100 MeV of the 3FGL Fermi
  catalogue \citep{fermi2015} for sources at high (low) galactic
  latitude\footnote{see
    \url{http://www.slac.stanford.edu/exp/glast/groups/canda/lat\_Performance.htm}
    for a plot of the sensitivity attained in four years.} is in fact
  $2\times10^{-9}$ cm$^{-2}$ s$^{-1}$ ($10^{-8}$ cm$^{-2}$ s$^{-1}$),
  assuming a power-law spectrum with index equal to 2. This
  corresponds to $\simeq1.5\times10^{-12}$ erg cm$^{-2}$ s$^{-1}$
  ($\simeq7.5\times10^{-12}$ erg cm$^{-2}$ s$^{-1}$).  A steady
  0.1-100 GeV flux like the one observed from {\1023} in the {\it
    sub-luminous} disk state, $F=(4.6\pm0.6)\times10^{-11}$ erg
  cm$^{-2}$ s$^{-1}$ \citep[see][and Sec.~\ref{sec:1023}]{takata2014},
  would have been observed by Fermi/LAT up to $\simeq7$ kpc (3.3 kpc
  if the source were at a lower latitude). On the other hand, a high-latitude
  low-mass X-ray binary like Cen X-4 is instead located at a similar
  distance than {\1023}, 1.2 kpc, and would haven then been easily
  seen by Fermi/LAT, while it has not been detected so far.

\subsection{Propeller scenario}

A scenario based on a propelling NS with an accretion rate of
$\mbox{few}\times10^{-11}$ M$_{\sun}$ yr$^{-1}$ naturally reproduces
the bolometric luminosity between few $\times 10^{34}$ erg s$^{-1}$,
observed from transitional pulsars like {\1023} and {\xss} in the disk
sub-luminous state (see Fig.~1 of this paper, and Fig.~2 of
\citealt{campana1998}).  We showed that we could reproduce the
  gamma-ray part of the SED as due to self-synchrotron Compton
  emission originated at the turbulent boundary between a propelling
  magnetosphere and the disk in-flow, assuming that there is a region
  where electrons can be accelerated to relativistic energies. The
  X-ray emission is instead due by the sum of synchrotron emission
  originated from the same region, and by the luminosity emitted by
  the accretion flow. For the accretion flow luminosity not to exceed the
  observed X-ray luminosity, an X-ray disk radiative efficiency of
  less than 20 per cent is requested.

The observed SED was reproduced by our modelling considering
  values of the fastness between 1.5 and 2.5, and assuming that 99 per
  cent of the disk flow is ejected by the NS, and that 15 per cent of
  the gravitational energy advected in the disk is available to power
  the propeller radiative emission. For the considered values of the
  fastness, the disk-magnetospheric boundary is highly magnetized
($\approx 10^{6}$ G; resulting from the dipolar contribution from the
pulsar) to produce a synchrotron emission cut-off at energies of about
1--10 MeV. If the acceleration region is small enough it can then give
a sizable SSC contribution, enough to explain she observed gamma-ray
emission. Our modelling indicates that a region with a volume of
$V \la 10^{15}$ cm$^{3}$ is needed, corresponding to a
sphere of radius equal to $\la 1$ km. Such a relatively small size
could be attained if the acceleration process takes place, e.g., in
filaments along the magnetic field lines at localized spots of the
disk-magnetospheric boundary.  As the volume of the emitting
  region decreases with decreasing $\omega_*$, we find the higher
  values of $\omega_*$ considered in this work (i.e. between 2 and
  2.5) more likely. For similar values of $\omega_*$, 3D MHD
  simulations performed by \citep{lii2014} showed that accretion and
  ejection of matter can coexist, with most of the matter being flung
  out by the fast rotating magnetosphere.

 The impossibility of observationally separating contributions of the
 accretion flow and the propeller just at the X-ray domain, is however
 partially limiting the model testing. This gives a larger phase space
 of plausible parameters for the accretion flow component, which can
 accommodate several different values of the fastness (see Table
 1). This translates into a range of possible mass in-flow rate,
 between 1 and $3\times10^{-11}$ M$_{\odot}$ yr$^{-1}$.  It is also
 true that the more direct, testable model predictions happen in a
 range of energies with no sensitive coverage (at the tens of MeV
 regime), or at timescales for which Fermi-LAT is not enough sensitive
 to track them (e.g., searches for correlation of gamma-ray and X-ray
 fluxes at the 100 s timescales cannot proceed). We also note that
 this model predicts no detectable TeV counterparts, what can be
 proven by extrapolating the predicted (or fitted) gamma-ray spectra
 to this domain and comparing with sensitivities of current or
 foreseen instruments \citep{actis2012}.
  
A Fermi process is a possible mechanism to accelerate electrons in the
magnetized, shocked environment expected at the boundary between the
disk and a propelling magnetosphere
\citep{bednarek2009,torres2012,papitto2012}. In P14, we assumed that a
first-order process injected energy in the electron distribution at a
rate:
\begin{equation}
\label{eq:acc}
\ell_{acc}=1.4\times10^5 \xi_{0.01}\;\bar{B}_6\:\mbox{erg s}^{-1},
\end{equation}
where $\xi_{0.01}$ is the acceleration parameter in units of 0.01, and
$\bar{B}_6$ is the strength of the magnetic field at the interface
$\bar{B}$, in units of $10^6$ G. In P14 we showed that the most
efficient radiative processes at the boundary between an accretion
disk and the propelling magnetosphere of a ms pulsar are synchrotron
interaction of the accelerated electrons with the NS field lines, and
Compton up-scattering of the synchrotron photons by the same
population of relativistic charges.  Synchrotron losses proceed at a
rate:
\begin{equation}
\ell_{syn}=1.1\times10^5\,\bar{B}_6^2\,\gamma_4^2\,\mbox{erg s}^{-1},
\end{equation}
where $\gamma_4$ is the maximum electron energy $\gamma_{max}$ in
units of $10^4$, while we assume SSC losses to be
$\ell_{SSC}=(f-1)\ell_{syn}$, where $f$ is implicitely defined as the
ratio between the total luminosity ($\ell_{syn}+\ell_{SSC}$) and the
synchrotron luminosity $\ell_{syn}$. Equating these radiative losses
to the energy input of the Fermi process of electron acceleration
(Eq.~\ref{eq:acc}) yields the maximum electron energy :
\begin{equation}
\label{eq:gammamax}
\gamma_{max}=8.2\times10^3\;f_{2}^{-1/2}\:\xi_{0.01}^{1/2}\;\bar{B}_6^{-1/2},
\end{equation}
where $f_2$ is the ratio between the total (synchrotron+SSC) and the
synchrotron luminosity $f$, in units of 2. The observed high-energy
cut-off of the observed spectral energy distribution indicates
$\gamma_{max}=10^4$, and the values of $f$ and $\bar{B}$ of our
modelling (see Table 1) indicates that the previous relation is
satisfied for an acceleration paramater $\xi$ ranging from $0.03$ to
$0.1$.

Reconnection of magnetic lines twisted in the turbulent region could
also provide accelerated particles. Such possibilities in the case of
a white dwarf propeller was studied by \cite{Meintjes2000} for the
case of AE Aqr.

According to our model, a relatively strong magnetic torque is needed
to power a propeller emission of $\simeq2.5\times10^{34}$ erg
s$^{-1}$, i.e.  of the order of that observed in X-rays and gamma-rays
from a system like {\1023} in the {\it sub-luminous} disk state. For
$\omega_*=2.5$, the torque expressed by Eq.~\ref{eq:torque}
corresponds to an expected spin-down rate, $\dot{\nu}=N_{mag}/2\pi
I\simeq-4\times10^{-15}$ Hz/s, where a NS moment of inertia
$I=10^{45}$ g cm$^2$ was assumed. Lower values of the fastness give a
larger expected spin down rate. If the system is in a propeller state,
a $\dot{\nu}$ larger by more than a factor of $2$ with respect to that
observed when the system is observed as a radio pulsar
($\dot{\nu}_{dip}=-1.9\times10^{-15}$ Hz s$^{-1}$, Archibald et
al. 2013), is then expected. The spin evolution during the disk
  state can be estimated by following the variations of the spin
  frequency measured from X-ray pulsations emitted in such a state,
  and/or by a comparison with the frequency of radio pulsations that
  will be observed when the rotation-powered pulsar will be back on.
Such a measure will then will then allow us to estimate the torque
acting onto the NS, and test our assumption that the system lies in a
propeller state when it has a disk.

The variability of the X-ray emission over time-scales of few tens of
seconds is a characteristic feature of millisecond pulsars in the disk
sub-luminous state. In the context of the propeller model we propose,
it can be attributed to variations of the mass-inflow rate (see
Eq.~\ref{eq:sys}), and the related response of the fastness and of the
location of the inner disk radius (Eq.~\ref{eq:rin}).  The observed
variability time-scales are indeed compatible with the viscous
time-scales in the inner parts of an accretion disk, as noted by
\citealt{patruno2014}. However, the lack of information on the
possible correlation between X-ray and gamma-ray emission on
time-scales of few hundreds of seconds prevents us to model the
variability of the SED, as it was done for the average SED.
\citet{linares2014} and \citet{patruno2014} argued that at the lowest
X-ray luminosity observed in the case of M28I and {\1023},
respectively, $L_X\simeq 5\times10^{32}$ erg s$^{-1}$, the inner disk
radius could expand beyond the co-rotation surface and a radio pulsar
turn on. Even in this case, the same considerations based on the
energy budget made in Sec.~5.2 suggest us to exclude a scenario in which the
observed gamma-rays owe only to intervals during which a
rotation-powered pulsar would be turned on.  \citet{tendulkar2014}
showed that {\1023} spends less than 25 per cent of the time in the
low/dipping state; if gamma-rays are only produced in this state a
0.1--100 GeV luminosity equal to four times the observed one (i.e.,
$\approx 4\times10^{34}$ erg s$^{-1}$) would be implied. This would
require conversion efficiency of spin down power into gamma-rays of 90
per cent, not taking into account the emission not falling into the
LAT energy band.

Also the bright, flat-spectrum radio emission observed from
transitional ms pulsars in the {\it sub-luminous} disk state
\citep{hill2011,deller2014} can be taken as an indication of
synchrotron emission originating from an outflow from the system. As
we already noted in P14, the properties of the emitting region where
we assumed that the high-energy emission originate are such that
synchrotron emission would be self-absorbed. The observed radio
emission should then be produced by a wider, optically thinner region
such as a compact jet.  

The necessity that a pulsar in a LMXB passes through a propeller
regime when the mass accretion rate decreases was already identified
since decades \citep{illarionov1975}, and reproduced by
magneto-hydrodynamical simulation \citep[][and references
  therein]{romanova2014}. However, observational evidences have been
few and indirect, among which the rapid decrease of the X-ray emission
at the end of X-ray outbursts of Aql X-1
\citep{campana1998b,campana2014}, accompanied by a hardening of the
X-ray spectral shape \citep{zhang1998}, is probably the most
remarkable.  Transitional ms pulsars demonstrated to be
exceptional laboratories to study not only the evolutionary link
between radio and X-ray millisecond pulsars, but also the centrifugal
inhibition of accretion. During its 2013 outburst {\m28} showed marked
flux and spectral variability that were interpreted by
\citet{ferrigno2014} as due to the onset of propeller reduction of
the mass in-flow. Furthermore, we showed how the {\it sub-luminous}
disk state of {\1023} can be naturally interpreted with a propeller
scenario, similar to the case of {\xss} \citep{papitto2014}. Future
observations will be fundamental to detect direct evidences of
out-flowing plasma, such as spectral lines and a possible correlation
between the radio and X-ray emission. Also, they will be used to
extend the knowledge of the SED and possibly
detect an excess of emission with respect to the spin-down power,
which would rule out a rotation-powered pulsar interpretation.

\section*{Acknowledgments}

We acknowledge support from the the grants AYA2012-39303 and SGR
2014-1073. AP is supported by a Juan de la Cierva fellowship. DFT
further acknowledges the National Natural Science Foundation of China
via NSFC-11473027 and the Chinese Academy of Sciences visiting
professorship program 2013-T2J0007.  We thank L. Burderi, E. de O\~na
Wilhelmi, T. Di Salvo, J. Li, N. Rea and M. Romanova for useful
discussion and comments. We acknowledge the International Space
Science Institute (ISSI) that funded an international team devoted to
the study of transitional ms pulsars where this work has been
discussed, and we thank all the members of the team for the fruitful
discussions. We thank the anonymous referee for his/her useful
comments and suggestions, as well as Jason Hessels and Amruta Jaodand
for their comments and for identifying a few typos in the first
version of the manuscript.

\label{lastpage}

\bibliography{biblio} 

\end{document}